\documentclass{aa}
\usepackage{txfonts}
\usepackage{graphicx}
\usepackage{natbib}
\bibpunct{(}{)}{;}{a}{}{,}
\usepackage{color}

\begin{document}

%\title{IR luminosity of galaxy clusters}
\title{The Infrared Luminosity of Galaxy Clusters}

%\subtitle{II. An example text with infinitesimal
%  scientific value\\
%  whose title and subtitle may also be split}

\author{M. Giard\inst{1}
  \and
  L. Montier\inst{1}
  \and 
  E. Pointecouteau \inst{1}
  \and
  E. Simmat \inst{1,2}}

\offprints{M. Giard, \email{ martin.giard@cesr.fr}}

\institute{ Centre d'Etude Spatiale des Rayonnements, CNRS/UniversitŽ\'e de Toulouse,
9 Avenue du colonel Roche, BP 44346, F-31028 Toulouse cedex 04, France
\and
Max-Planck-Institut fuer Kernphysik, 
Saupfercheckweg 1, D-69117 Heidelberg, Germany}

\date{Sent 23 june 2008 / Accepted 31 july 2008}

\abstract
   {The cosmological models for the formation of the first stars and the large scale structures now raise the question of how many dust particles were released to the general diffuse gas and how these impact the star formation process. In this framework, we focus on the scale of galaxy clusters.}
   {The aim of this study is to quantify the infrared luminosity of clusters as a function of redshift and compare this with the X-ray luminosity. This can potentially constrain the origin of the infrared emission to be intracluster dust and/or dust heated by star formation in the cluster galaxies.}
   {We perform a statistical analysis of a large sample of galaxy clusters selected from existing databases and catalogues.We coadd the infrared IRAS and X-ray RASS images in the direction of the selected clusters within successive redshift intervals up to $z = 1$.}
   {We find that the total infrared luminosity is very high and on average 20 times higher than the X-ray luminosity. 
   If all the infrared luminosity is to be attributed to emission from diffuse intracluster dust, then the IR to X-ray ratio implies a dust-to-gas mass abundance of $5 \times 10^{-4}$. 
   However, the infrared luminosity shows a strong enhancement for $0.1 < z < 1$, which cannot be attributed to cluster selection effects. 
   We show that this enhancement is compatible  with a star formation rate (SFR) in the member galaxies that is typical of the central Mpc of the Coma cluster at $z = 0$ and evolves with the redshift as $(1+z)^5$. }
{It is likely that most of the infrared luminosity that we measure is generated by the ongoing star formation in the member galaxies. From theoretical predictions calibrated on extinction measurements (dust mass abundance equal to $10^{-5}$), we expect only a minor contribution, of a few percent, from intracluster dust.}


\maketitle

\section{Introduction \label{intro}}
Dust is only a minor constituent of the known cosmic matter. 
It represents only 1\% of the total diffuse (non-stellar) baryonic mass in our Milky Way
(see e.g. \citet{Spitzer78}, page 7 and 8), and possibly 0.002\% in the
diffuse intergalactic medium \citep{Aguirre01}.
However,  small solid bodies (i.e. dust grains) are the main source of opacity for the
electromagnetic radiation emitted by stars because they allow some electrons to be far more mobile than
in free atomic or molecular orbitals.  The practical consequence
of this opacity is that dust thermal re-emission of absorbed
photons has become a very efficient and popular tracer of the star
formation efficiency (see the review by \citet{Lagache05} and references therein).

Moreover, the photodetachment of electrons from dust grains can occur more easily 
than from either a free atom or molecule, because the binding
energies are smaller. This implies that the gas in which dust grains
are embedded can be heated by photoelectrons if the stellar radiation
is sufficiently strong (see \citet{Weingartner06}
and references therein).  Heating by the photoelectric effect of
grains and cooling by their thermal infrared (IR hereafter) emission,
are two opposite energy routes that, in addition to the gas
processes, can have an important impact on the energy balance of gas
clouds, and therefore on their ability to fragment, condense and form stars
\citep{deJong77}.  Finally, it has been conjectured
  since the advent of molecular astrophysics \citep{Gould63} that the grains surfaces play an important
  role in interstellar chemistry, and particularly in the synthesis of
  molecular hydrogen (see \citet{Cazaux02} for a
  recent estimation).

The question of whether dust could play a role in optically thinner
  media and on cosmological scales such as the intergalactic medium
  (hereafter IGM) or the intracluster gas
  (hereafter ICM), if present, has  been scrutinized on a theoretical 
  point of view by several authors: see \citet{Dwek90}, \citet{Popescu00}, and \citet{Montier04} . 
\citet{Montier04} computed the balance of dust heating and cooling with respect to the
  dust abundance: cooling dominates at high temperatures in the hot
  gas of virialized structures (i.e. galaxy clusters), and heating
  dominates in low temperature plasma under high radiation fluxes such
  as in the proximity of quasars. The details, of course, depend on
  the local physical parameters such as the grain size and gas
  density.  The limitations of and uncertainties in the modeling are due to incomplete 
  knowledge how both the formation of the first supernovae and the production of dust, 
  correlate spatially and precede in time the heating of gas accreted onto the large-scale structures and dust destruction.
The processes of feeding and enriching both the ICM and IGM with dust
is even more of a key issue when taking into account that dust in
these media is affected strongly by thermal sputtering and has
lifetimes ranging from $~10^6$~yr to $~10^9$~yr (depending on the
grain size).  
Therefore, when attempting to understand in detail
the formation of large-scale structures in the Universe, the impact
of a dust component on the thermodynamics of the early ICM raises the
important question of the heavy elements and dust grain synthesis in
the first supernovae (SNe) and how they were able to seed the primordial
diffuse gas and change its ability to fragment and
form more stars and galaxies (see for instance \citet{Schneider06} or 
\citet{Kampakoglou08}).

\citet{Elfgren04} and \citet{Elfgren07} evaluated the dust emission in large scale
structures from the time of reionization to the present by 
assuming the dust density to be a fraction of the
  dark matter density and its mass distribution to follow that of the
  dark matter, and by fixing the dust lifetime. Depending on the
dust lifetime, these authors found that the integrated signal from the dust thermal emission
produces brightness fluctuations on arcminute scales, which can be
comparable to CMB and galactic dust fluctuations.

We therefore expect dust in the IGM/ICM to contribute to the diffuse IR emission 
in the direction of the large scale structures, and particularly towards galaxy clusters. 
The only claim of detection of IR emission from intracluster dust was that of
\citet{Stickel98,Stickel02} towards the
Coma cluster using the ISOPHOT instrument. A statistical detection of intracluster
dust extinction towards SDSS quasars located behind clusters of
galaxies was obtained by \citet{Chelouche07}. 
An upper limit was also derived by \citet{Muller08} by studying the statistical reddening of galaxies behind 458 RCS clusters.
Finally, \citet{Bai07b} derived
upper limits at 24 and 70 $\mu$m from observations by the
Spitzer telescope of the cluster Abell 2029 ($z = 0.08$). Detecting
emission from the ICM is indeed a difficult task because it must be distinguished from the
IR emission of the cluster galaxies. The overall cluster IR emission for a
single cluster is comparable or lower than the average sky fluctuation
produced by galactic cirrus clouds and the cosmic infrared background
(due to background galaxies).

From observational results, the IR emission from clusters is
expected to originate primarily in star-forming cluster galaxies. 
\citet{Montier05} performed a statistical detection of
this overall dust emission in the direction of clusters  by
stacking the IRAS emission at 12, 25, 60, and 100 $\mu$m towards more
than 11~000 known galaxy clusters. They measured
significant, but faint, emission at all IRAS wavelengths. This
emission corresponded to the dust emission originating in both intracluster
dust and dusty star-forming cluster galaxies.

From measurements of the IR emission of star-forming
  galaxies from observations by the Spitzer telescope at 24 $\mu$m, \citet{Bai06,Bai07}
  derived the luminosity function of infrared galaxies in the Coma
  cluster ($z=0.02$) and MS1054-0321 ($z=0.83$). They demonstrated that
  the cluster galaxy luminosity function resembles that of
  field galaxies at both studied redshifts. They also emphasized the strong
increase in IR luminosity with redshift, which implied that 
the star formation rate (SFR) was higher in the
past within cluster environments as it was
for field galaxies. Their luminosity functions, when
integrated over the entire range of galaxy masses, inferred a total
infrared bolometric luminosity of $2.15\times 10^{45}\, h_{70}^{-2}$
and $27\times 10^{45}\, h_{70}^{-2}$~erg/s respectively for Coma and
MS1054-0321 (\citet{Bai07} and private communication from
the author).  In contrast, the rest-frame X-ray bolometric
luminosities of the two clusters are respectively $1.23\times 10^{45}\, h_{70}^{-2}$~erg/s
\citep{Arnaud99} and $3.05\times 10^{45}\,
h_{70}^{-2}$~erg/s \citep{Kotov05}.

In this paper we use the stacking method formerly developed by \citet{Montier05} to probe statistically the IR dust
luminosity of galaxy clusters as a function of redshift. Our
objective is to determine whether this luminosity evolves with redshift and discriminate
if possible between the contribution of the SFR
in member galaxies and the diffuse emission from intracluster dust.  
The stacked analysis was also completed on the X-ray Rosat All Sky Survey (RASS). 
We used the RASS as a reference, because the X-ray properties of galaxy clusters (i.e. structural and scaling properties) are well studied and constrained (see reviews by \citet{Voit05} and \citet{Arnaud05}). More specifically, it is well known that the X-ray luminosity correlates well with the temperature or mass of clusters (\citet{Arnaud99}, \citet{Maughan07}, \citet{Vikhlinin08} or \citet{Pratt08}).

The paper is organized as follows. Sections \ref{datasets} and
\ref{stacking} present the details of the data used, the selection of
the fields stacked, and the stacking method. Our results are presented and
discussed in Sects. \ref{results} and \ref{discussion}.

In this paper, we use the following cosmology: $H_0 = 70 km/s/Mpc$, $\Omega_m = 0.3$ and $\Omega_\Lambda = 0.7$.

\section{Data sets and catalogues \label{datasets}}

\subsection{The catalogues of clusters \label{catcl}}

\begin{table}[t!]
\begin{center}
\begin{tabular}{|l|c|c|c|}
\hline
Numbers in each Catalogue & NED & NSC & SDSS\\
\hline
All references & 18~829 & 8~155 & 13~823\\
Distinct fields (10\arcmin) & 13~746 & 7~888 & 11~982\\
\hline
\underline{Fields rejected :} & & &\\
%\cline{1-1}
Data missing & 1~665 & 972 & 1~545\\
Gal. plane: $|b_{II}| < 10^{\circ}  $ & 17 & 0 & 0\\
Strong sources & 1~651 & 947 & 1~063\\
Noisy fields & 2~120 & 1~244 & 1~898\\
\hline
Number of fields stacked & 8~293 & 4~725 & 7~476\\ 
\hline
\end{tabular}
\caption{Selection of the cluster fields selected for stacking.}
\end{center}
\end{table}

\begin{figure}[t!]
\includegraphics[width=9cm]{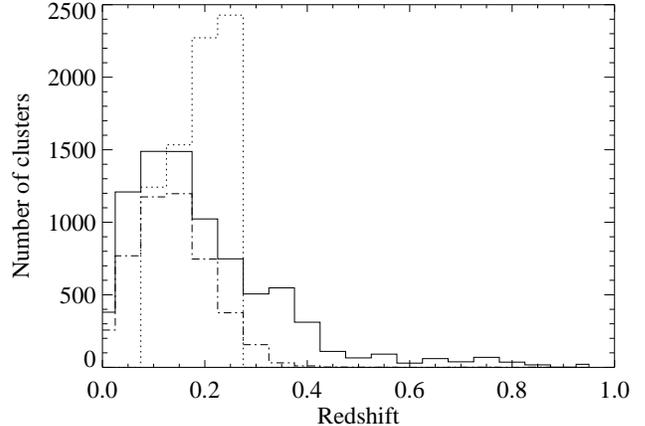}
\caption{\label{Fig_histo_cat} Distribution in redshift of the
  stacked clusters for the three selected lists: NED (solid line), NSC
  (dashed-dotted libe), and SDSS (dotted line). See Sect.~\ref{catcl}
  for the definition of the cluster lists.}
\end{figure}

We used three different lists or catalogues of galaxy clusters as
input to our stacking method.
\begin{itemize}
\item The first one was a generic extraction from the NED
  database\footnote{see the NED web site:
    http://nedwww.ipac.caltech.edu/index.html} of objects
  classified as ``clusters of galaxies''. Each object has a
  coordinate and redshift. This list is extremely inhomogenous. 
  We started with a total of 18~829 references, which reduced to 13,746 clusters fields 
  after source duplication and overlapping fields had been identified (see
  Sect. \ref{stacking}). Hereafter, we refer to this list as the NED
  sample. This is the same as the cluster list used in our previous investigation \citep{Montier05}.

\item The second selection corresponds to the Northern Sky optical
  Cluster survey (NSC hereafter) by \citet{Gal03}. It contains 8~155 references,  each with an indicator
  of richness, $N_{gals}$.

\item The third catalogue is the SDSS/MaxBCG cluster catalogue (SDSS hereafter) from
  \citet{Koester07}, which provides a precise measurement
  of the cluster richness ($N_{200}$, i.e. the number of galaxies
  within $R_{200}$). However, the redshift coverage of this selection
  is limited to $[0.1,0.3]$. We begin with 13~823 references.

\end{itemize}

In Fig. \ref{Fig_histo_cat} we provide the redshift histograms of three sets of stacked fields (see the description of the selection
process in Sect. \ref{stacking}).

\begin{figure*}[!th]

\includegraphics[width=9cm]{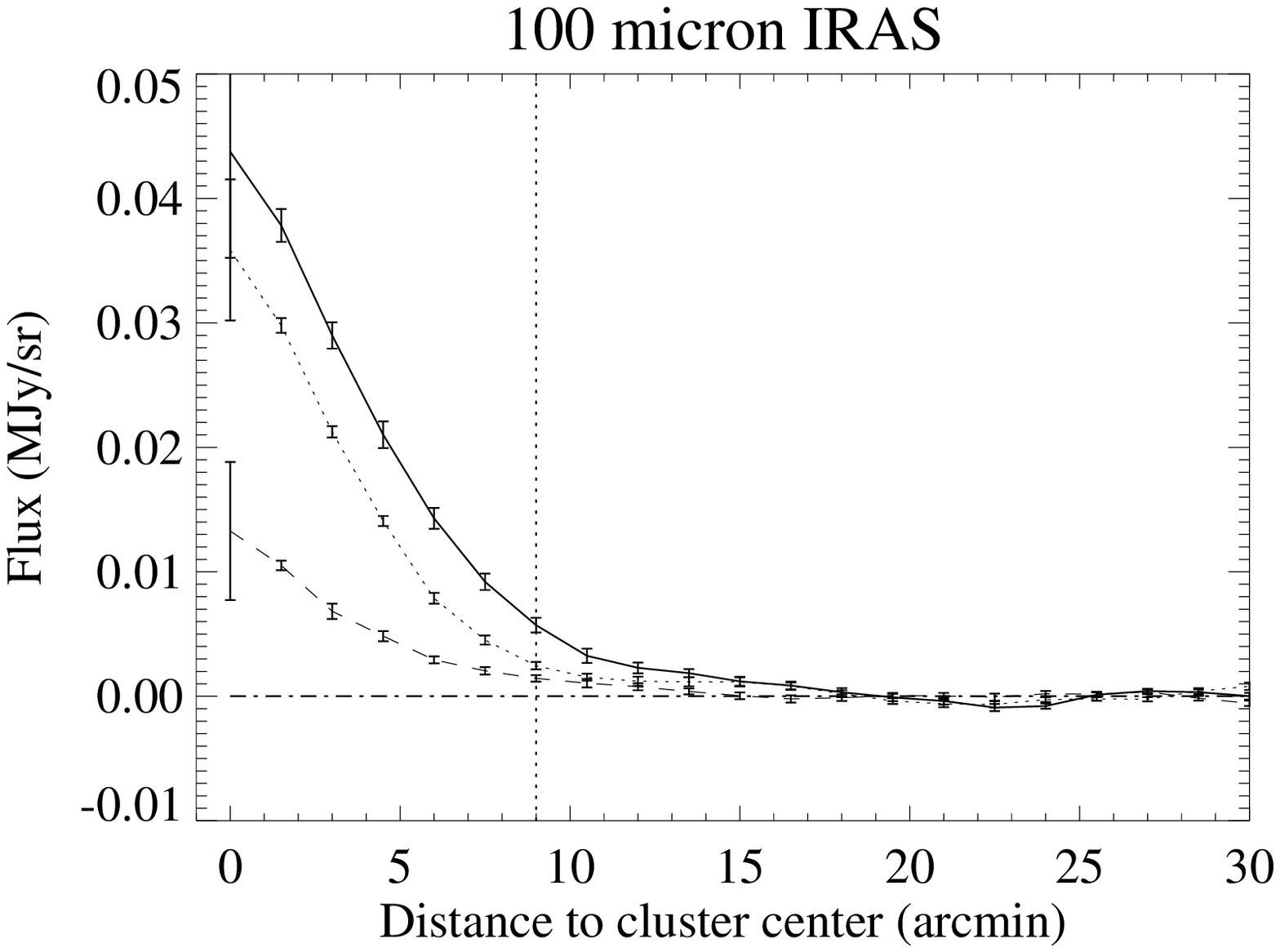}
\includegraphics[width=9cm]{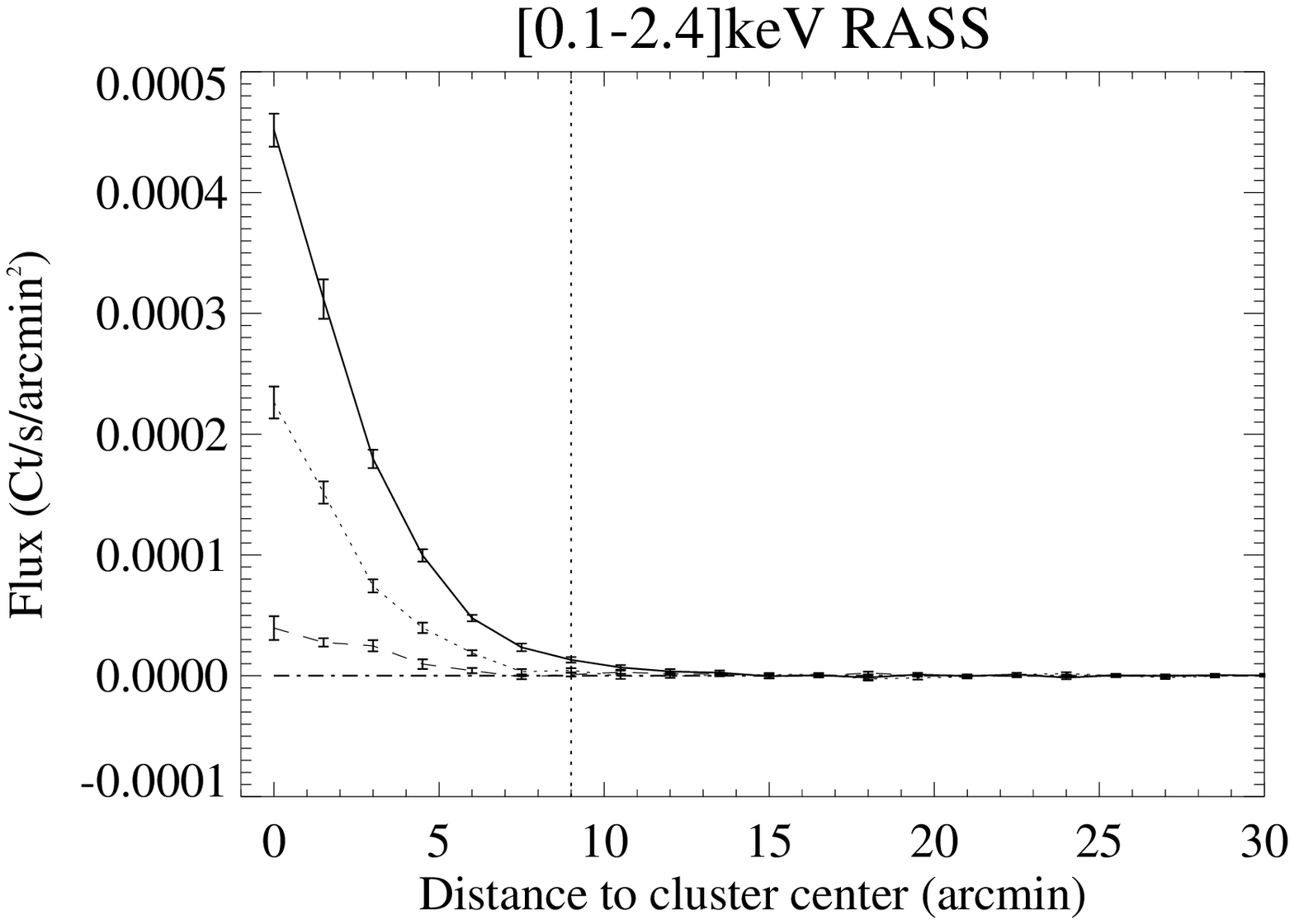}

\caption{ \label{Fig_profiles} Radial profiles derived from the 
  stacked images of the NED selection in three redshift bins centered at
  $z=0.09$ (solid line), $z=0.24$ (dotted line) and $z=0.62$ (dashed
  line) for the 100 $\mu$m IRAS survey (left panel) and the
  [0.1-2.4]keV RASS (right panel). The redshift bins were defined to each
  have equal numbers of galaxy clusters (about 830 -- see
  text). The dashed vertical line shows the limit of the aperture used
  to compute the luminosities and the dotted-dashed horizontal line is
  the zero flux level.
}
\end{figure*}

\subsection{All sky survey data sets}

We performed our stacking analysis over five sets of all-sky survey
data: the IRAS at 12, 25, 60, and 100~$\mu$m data and the ROSAT All
Sky Survey broad band data (i.e. [0.1-2.4]keV RASS).  For the IRAS
survey, we used the reprocessed IRIS data set that was
recalibrated with DIRBE and cleaned of the instrumental stripes \citep{Miville05}.

For the RASS, we first derived count rate images that where computed to be the ratio
of the count images (i.e. rs93*im1.fits) to the corresponding exposure
map (i.e. rs93*mex.fits\footnote{see the ROSAT Status Report \#180 (ROSAT NEWS No. 70) on the
  web  for a detailed description of the
  RASS products: http://heasarc.nasa.gov/listserv/rosnews/msg00126.html}). 
  We then converted these count rate images into
unabsorbed flux images by correcting for the absorption due to the
hydrogen column density in the [0.1-2.4]keV band. We used a fixed
conversion factor from counts to fluxes of $1.5\times 10^{-11}
\textrm{erg cm}^{-2}$ for one count. This factor was derived using a
\texttt{WABS*MEKAL} model with XSPEC (see \citet{arnaud96}) assuming a
temperature of 1.5~keV and a metallicity of $Z=0.3$~Z$_\odot$ for the
intracluster hot plasma at the median redshift of $z=0.2$ (see
Fig~\ref{Fig_histo_cat}), and a hydrogen column density of $3 \times
10^{20} \textrm{cm}^{-2}$, which was the average for the fields stacked. 

The value of the conversion factor was
almost insensitive to the cluster redshift. From $z=0.01$, there is a
slight increase of $\sim3$\% up to $z=0.4$ and of $\sim 10$\% up to
$z=1$. 
However, this is quite sensitive to both the intracluster gas temperature and the column density of the
foreground HI. 
By varying the intracluster gas temperature from 1 keV to 5 keV, 
the conversion factor is changed from $1.4 \times 10^{-11} \textrm{erg cm}^{-2}$
to $1.65 \times 10^{-11} \textrm{erg cm}^{-2}$, respectively.
The most important effect originates in the foreground HI column density: the conversion factor rises from 
$1.2 \times 10^{-11} \textrm{erg cm}^{-2}$ to $1.8 \times
10^{-11} \textrm{erg cm}^{-2}$  for $N_{HI}$ that increases from  $1.5 \times 10^{20} \textrm{cm}^{-2}$ to $6
\times 10^{20} \textrm{cm}^{-2}$. 

Considering that the average HI column density of the fields stacked is $3 \times
10^{20} \textrm{cm}^{-2}$ with an rms fluctuation of $1.5 \times
10^{20} \textrm{cm}^{-2}$, we became concerned about the errors introduced into 
the stacked fluxes when using a single conversion factor.
We completed the stacking analysis therefore using a conversion factor for each cluster corresponding
to its individual $N_{HI}$ value. Surprisingly there was almost no detectable impact on the averaged fluxes obtained after stacking.
The reason was that, although the conversion factor varies significantly from cluster to cluster, the 
averaging process is completed for such a high number of clusters that the effect of any scatter on the averaged image diminishes to zero.
The same was true for the cluster gas temperature, for which an initial scattering in the conversion factor was smaller. 
The average temperature used, 1.5~keV, was consistent with the average richness of the clusters that we selected from the
SDSS entries, $<N_{200}>~=~16$,  and the $N_{200}$ to gas temperature correlation  derived by \citet{Rykoff08} for the same  clusters (see their Table~4). 

After calculating the stacked fluxes, they were converted to luminosities in the [0.1-2.4]keV band using  the associated luminosity distances computed for the adopted cosmology (see Sect.~\ref{intro}). 

Finally, we derived the X-ray bolometric luminosities from the broad band [0.1-2.4]keV luminosities using a correction factor of 1.35,  which was also computed by XSPEC using a \texttt{MEKAL} model for a cluster at $z=0.2$ of temperature $kT=1.5$~keV and 
metallicity $Z=0.3$~Z$_\odot$). The bolometric correction was mostly sensitive to the cluster gas temperature with a conversion factor of 1.24 at 1~keV (or 2.07 at 5~keV). However, the significant size of our cluster sample enables us to use a single correction factor computed for the average gas temperature.

We were able to confirm our luminosities by comparison with those given by \citet{Bohringer04} in the same energy band  for clusters in the REFLEX catalogue; these numbered about one hundred of clusters with luminosities ranging from a few $10^{42}$ erg/s to a few  $10^{45}$ erg/s. Our luminosities were systematically slightly lower than the REFLEX ones with a mean difference of $ -16\%$ and a  rms scatter of $\pm  13\%$.

\section{Stacking the IR and X-ray maps in different redshit bins \label{stacking}}

The entire stacking method used was similar to that described by
\citet{Montier05}. We recall the main steps of the analysis:

\begin{enumerate}

\item {\bf \emph{Distinct fields}} :ÊWe first eliminated multiple occurrences of the same
  field. This could occur for two reasons. First, because several
  clusters were neighbors on the sky (at the chosen
  scale of our fields). Second, because the lists of clusters that we used,
  as in the case of the NED list, contained several references to the
  same object. In practice, once a field had been selected for
  stacking, it was identified by the coordinates of its center and any
  other field whose center coordinates were closer than 10 arcmin was not selected.

\item{\bf \emph{Data missing}} : We then retained only fields for which data existed in all
  five sky surveys (i.e. the IRIS 12, 25, 60, and 100 $\mu$m surveys and the
  RASS). This step eliminated 1~665 fields from the NED list, 972 from the
  NSC catalogue, and 1~545 from the SDSS catalogue. At this stage, we
  extracted a $1^\circ\times 1^\circ$ sq. degree field for each of the
  relevant cluster direction. The pixel size of all extracted field was
  set to IRIS pixel size of 1.5 arc-minutes. The RASS images had an
  original pixel size of 0.75 arc-minute, which was rebinned to 1.5
  arc-minutes by coaddition of pixels and nearest neighbor
  approximation. This rebinning was performed to facilitate the data
  handling process for the different surveys.

\item {\bf \emph{Galactic plane}} : We excluded all fields within a distance of 10 degrees from the
  Galactic plane (NED references).

\item {\bf \emph{Strong sources}} : In each field, we then masked all obvious sources in a
  $5\times 5$ pixels square area. The sources were first taken from the
  IRAS point source catalogue. This was completed by a source detection
  algorithm for both the IRIS and the X-ray images. This algorithm
  performed a 3x3 boxcar smoothing of each field and selected
  sources identified at a confidence level of 5 sigma with respect to the average. If
  an infrared source was detected at the center of a field (i.e.
  within a radius of 5 pixels) the field was removed from the
  list. However, we allowed fields to have an X-ray source detected at
  their center since massive galaxy clusters were well detected in
  the RASS data. This eliminated 1~651 more fields from the NED
  list, 947 from the NSC, and 1~063 from the SDSS lists.

\item  {\bf \emph{Background subtraction}} :  
At this stage, we performed a background subtraction for each
  map. We chose a 3 degree polynomial surface as background layer. 
The best polynomial was obtained by applying a least square fit to the image 
over an external ring which extended
  from pixel \# 10 to pixel \# 20 (i.e. between radii of 15' to 30' from center). The
  source mask defined previously in each sky direction was used
  to blank the image area used for the fit. The order chosen for the polynomial surface 
  ($3^{rd}$ order) is optimum because it is able to adapt to
  most backgrounds without altering data at the field center due to some uncontrolled oscillation in the fit.

\item  {\bf \emph{Noisy fields and weighting}} :
Finally, we eliminated the noisiest fields and weighted the remaining fields with their intrinsic noise level at $100~\mu$m  wavelength. 
This was performed  by computing the rms flux in each masked image over the outer ring defined for the background fit (between radii of  15' to 30' from center).  All fields for which this rms was
 higher than 5 times the standard deviation of all rms were then excluded.
This rms evaluation of the sky fluctuation performed on each field at 
$100~\mu$m was also used to compute the weight of each image in the
final stacking.  For field$\#i$ we used the following weighting:
$ \sigma_i = (rms_i+median(rms_i))^{-2} $. The median of all rms values
was added to each rms value to prevent any extremely flat images 
(which statistically happen) dominating the final sum.
This final selection step eliminated 2~120 more fields from the NED,
1~244 from the NSC, and 1~898 from the SDSS lists.
\end{enumerate}

Our final data set contained 8~293 fields from the NED list, and respectively 4~725 and 7~476 from the NSC and the SDSS catalogues. The average Galactic HI column density in the direction of the selected NED clusters was about $2.8 \times 10^{20}~ \textrm{cm}^{-2}$  (and $2.4 \times 10^{20}~ \textrm{cm}^{-2}$  for the NSC and SDSS selection). 

Regarding  the cluster richness, the average number of galaxies in the NSC and SDSS clusters selected were respectively $N_{gals} = 32$ and $N_{200} = 16$. From the correlation  between $N_{200}$ and the gas temperature derived for the SDSS  cluster catalogue by \citet{Rykoff08} (see their Table~4), we obtained a mean temperature of 1.5~keV for our SDSS selection. 

To study the evolution of the cluster luminosity with redshift, the selected cluster fields were sorted in order of increasing redshifts. Redshift bins were then defined to have an equal number of clusters in each bin (that is 830 clusters in each of the 10 bins for the NED selection).  The average flux image for each bin was calculated using the optimal weights specified above. The bins, weights, and source masks used in the stacking were the same at all wavelengths.

\begin{figure}[!t]
\resizebox{\hsize}{!}{\includegraphics{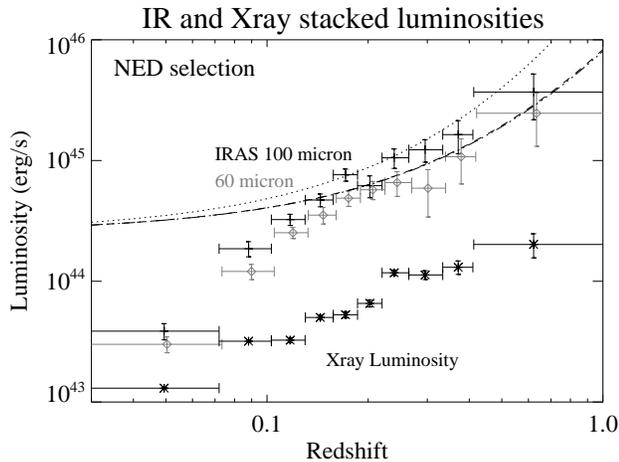}}
\caption{\label{Fig_luminozities} Averaged stacked luminosities into
  redshift bins for the NED sample. Crosses are the IRIS $100~\mu$m
  $\lambda I_{\lambda}$ luminosities and empty diamonds the IRIS
  $60~\mu$m ones. Stars correspond to the X-ray bolometric luminosities
  derived from the [0.1-2.4]keV RASS. The two curves show the total IR
  luminosity density ($erg/s/Mpc^2$) from the cluster galaxies
  computed with the SFR evolution law by \citet{Lefloch05} (dashed line) and by \citet{Bai07} (dotted line).
  Both curves are normalized on the central luminosity density of the galaxy clusters Coma and MS 1054 as done in \citet{Bai07}.}
\end{figure}

%For the NSC and SDSS sample we can also independently stack the clusters into bins of cluster richness to check the correlation between the richness and the luminosity at infrared and X-ray energies.

When a positive signal was detected at the center of a stacked field at a given redshift, we computed the corresponding averaged luminosity. This luminosity was calculated by integrating the flux inside the central 18' (9' radius from center) of the stacked images after subtraction of a possible residual background computed to be the average flux in an outer annulus between  9' and 18' radii from the field center. 

For the infrared data, the luminosity was computed for a wavelength interval of $\Delta \lambda = \lambda$. This is a convenient quantity since astrophysical dust models show that the total infrared luminosity can be well approximated as the sum of these contributions in the four successive IRAS bands \citep{Desert90}.

\section{Results \label{results}}
%\section{Origin of the  IR luminosity}

The stacking provided firm detections at all
redshifts only in the 60 $\mu$m, 100 $\mu$m and X-ray bands.

Fig. \ref{Fig_profiles} shows the radial profiles computed for the
stacked images of the NED selection at $100~\mu$m and [0.1-2.4]keV for three
redshift bins: $ z = 0.09$, $z = 0.24$, and $z = 0.62$. There is a
clear signal detected at all redshifts,  at both infrared and X-ray
energies.

In Fig. \ref{Fig_luminozities}, we show the average infrared and X-ray luminosities in each bin for the NED clusters sample. 

\begin{figure}[!t]
\resizebox{\hsize}{!}{\includegraphics{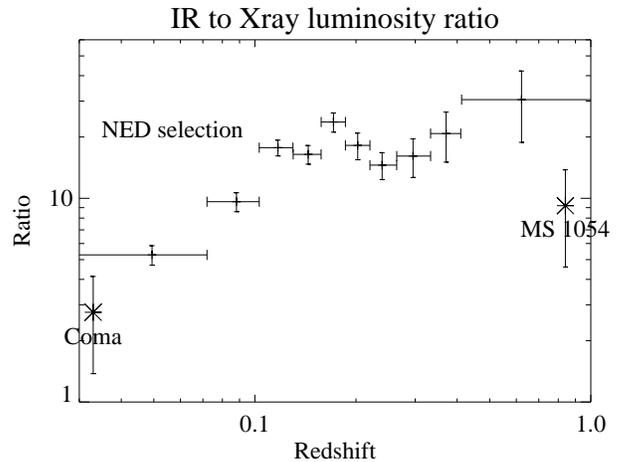}}
\caption{\label{Fig_IR2X} Ratio of the bolometric IR to X-ray
  luminosities in each redshift bin. The two stars indicate the IR-SFR
 to X-ray luminosity ratio for
  two cluster as derived by \citet{Bai07} from Spitzer
  observations (Note that for these clusters the IR luminosity is the integrated contribution from all infrared galaxies detected at $24 \mu m$). }
\end{figure}

In Fig. \ref{Fig_IR2X}, we show the ratio of the total infrared luminosity to the X-ray bolometric luminosity. According to the dust model of \citet{Desert90}, the total infrared luminosity (i.e. bolometric) was approximated by the sum of the 60 and $100~\mu$m luminosity : $L_{IR} \simeq \lambda I_{\lambda}(60~\mu m) + \lambda I_{\lambda}(100~\mu m)$. This value should certainly be taken to be a lower limit for the total IR luminosity since we know from direct observation or from modelling that dust emits a large fraction of its luminosity longward of $100~\mu m$, both in normal quiescent spiral galaxies \citep{Sauvage05} and for intracluster physical conditions \citep{Popescu00}. The two star symbols in the same plot indicate the IR-SFR to X-ray luminosity ratio derived by \citet{Bai07} for two peculiar clusters, Coma and MS 1054. The infrared luminosity of these two clusters was derived from Spitzer $24~\mu$m photometry of the member galaxies and does not include the diffuse emission if any is present. We assigned a $50 \%$ error bar to the IR luminosity of these two clusters as measured by the authors.

\subsection{Origin of the  IR luminosity}

The first conclusion to draw from our results is that the average bolometric infrared luminosity of galaxy clusters is far higher than their X-ray luminosity. The average IR to X-ray ratio is about 20, possibly showing a slight increase with redshift (see Fig. \ref{Fig_IR2X}).

Our measurements of IR luminosity probe the contributions of all 
sources of infrared emission in clusters; i.e. star formation in the member 
galaxies and any diffuse emission from intracluster dust.
These two emission processes have 
different physical origins and consequences. 
On one hand, infrared emission from
galaxies is correlated with the total star formation rate, which provides a net energy contribution to the immediate environment. 
On the other hand, emission from intracluster dust is a cooling agent that can accelerate 
 the condensation of diffuse baryons into stars and galaxies.

High IR to X-ray luminosity ratios were also measured for two individual clusters 
of galaxies by \citet{Bai07}:  $L_{IR} / L_X \simeq 2$ and $9$, for
Coma ($z=0.023$) and MS1054-0321 ($z=0.826$) respectively. 
However, in these two cases, the IR luminosity is due only to the
contributions of the individual cluster galaxies detected by the
Spitzer-IRAC camera at $24 \mu$m, and does not measure any diffuse
intracluster dust emission.

Although this comparison relies on only two clusters, and should therefore be considered with care,
it implies that the IR luminosity contribution of cluster member galaxies represents a significant fraction 
of the IR luminosity measured in our stacked data, although possibly not all of the luminosity. 
Any additional IR emission could either be due to a population of unresolved galaxies
in the Spitzer observations, such as IR emitting dwarf galaxies with low surface
brightness \citep{popescu02,hinz06}, or
to diffuse intracluster dust. 
In the following, we attempt to assess quantitatively the validity of
these two hypotheses: the IR originates in (1) cluster galaxies, or (2) intracluster dust. 

We note that both Coma and MS1054-0321 are well known
non-relaxed clusters that show significant evidence of ongoing dynamical activities
\citep{neumann03,gioia04}.

\subsection{Intracluster dust \label{ICdust}}

We use the simple model of intracluster dust emission developed by
\citet{Montier04}, where the dust grains are heated by
collisions with free electrons and radiate in the infrared with
properties typical of interstellar dust grains. To first
order, the IR to X-ray luminosity ratio is proportional to $Z_d = M_{dust}/M_{gas}$, which is the dust-to-gas mass ratio.
If we normalize on the curve shown in Fig. 8 of \citet{Montier04}, the IR to X-ray luminosity ratio over the relevant electron temperature range ($10^7$ to $10^8 keV$) is then given by: 

\begin{equation}
\frac{L_{IR}^{ICM}}{L_X^{ICM}} \simeq 0.40 \left( \frac{Z_d}{10^{-5}}
\right)
\label{IR2X}
\end{equation}

To derive an IR to X-ray ratio similar to that measured, i.e. 
$L_{IR} / L_X \simeq 20$, from the thermal emission of diffuse
intracluster dust alone, we therefore require a dust-to-gas mass ratio of $5~\times~10^{-4}$.

This dust-to-gas mass ratio would represent a significant fraction ($\sim
17$\%) of the total mass of metals available in an intracluster environment
(assuming $Z = 0.3~Z_{\odot}$). The precise dust-to-gas mass
ratio in the intracluster medium cannot be easily constrained on purely
theoretical grounds, since it is the result of a complex balance between
the production and destruction processes of ICM dust (see the
discussion in \citet{Montier04}).

\subsection{SFR of the member galaxies}

It is now well established that the infrared emission of galaxies
increases significantly between redshift $z = 0$ and $z=1$, because
the star formation rate (SFR) is measured to be higher in the past than the present. 
With Spitzer $24~\mu m$ data, the evolution of the infrared luminosity function of 
galaxies was characterized quantitatively for both field galaxies in the Chandra
Deep Field-South by  \citet{Lefloch05} and galaxies in massive clusters by \citet{Bai07}.

We adopt the Schechter formalism \citep{Schechter76} to count the
number of objects within each logarithmic (base ten) luminosity interval:
\begin{equation}
\Phi_{IR}(L) = \Phi_{IR}^{*} (L_{IR}/L_{IR}^{*})^{(1-\alpha)} \exp (-L_{IR}/L_{IR}^{*}) 
\end{equation}

\noindent where $\alpha  = 1.41 $ is fixed as in \citet{Bai06}.

Both \citet{Lefloch05} and \citet{Bai07} were
able to describe the evolution of this luminosity function using power
laws for $L_{IR}^{*}$ and $\Phi_{IR}^{*}$:
\begin{eqnarray}
L_{IR}^{*}  &  \alpha  & (1+z)^{\alpha_L} \nonumber \\
\Phi_{IR}^{*}  & \alpha & (1+z)^{\alpha_D}
\end{eqnarray}

\noindent with $\alpha_L=3.2$, $\alpha_D=0.7$ in \citet{Lefloch05},
or $\alpha_L=4.0$, $\alpha_D=1.4$ in \citet{Bai07}.
It is important to note that the two sets of coefficients
are compatible within the reported error bars in these two works.

We integrated the luminosity functions to derive the evolution of the total SFR luminosity with redshift.
The normalisation used at $z = 0$ is adjusted from
the so-called BvdHC sample of IR galaxies in the Coma cluster (see
filled circles in Fig. 5 of \citet{Bai07}) :
\begin{eqnarray}
\Phi(10^{41} erg/s, z=0) & =  & 59 / Mpc^2/\log_{10}(L_{IR}) \nonumber \\
L_{IR}^{*} (z = 0)  & = & 10^{10.49}\; L_{\odot} \nonumber \\
 & = & 1.18 \times 10^{44}\; {\rm erg/s}
\end{eqnarray}

The resulting total infrared SFR luminosities are plotted in
Fig.~\ref{Fig_luminozities} as dashed \citep{Lefloch05} and dotted \citep{Bai07} lines. These
integrated luminosities are actually close to a $(1+z)^5$
evolution law.  We emphasize that this normalisation of the SFR luminosity  is relative to a projected
cluster area of $1 Mpc^2$; it is a coincidence that it corresponds to the same level as the total luminosities computed from our stacked IRAS images at redshifts higher than 0.1. 
Although we do not exploit this coincidence, we believed that it was noteworthy for future quantitative investigations of this question.

\section{Discussion \label{discussion}}

A clear interpretation of our results is difficult because the  galaxy cluster sample
that  we extracted from the NED catalogue is obviously affected by  selection effects. 
For instance, the net decrease in
the IR luminosities compared with the SFR model at redshifts lower than
0.1 is likely to be a strong selection effect in the cluster mass
function that produces a strong bias towards low mass clusters at low
redshifts with respect to high mass clusters (i.e. the relative number of identified low
mass with respect to high mass systems  is far more important at low than at high
redshift).  This issue was the first motivation for stacking the cluster data
at X-ray wavelengths and to use the X-ray data as a reference data set. In a
second time, to tackle this problem further, we repeated our
stack analysis for the NSC catalogue, which has a more carefully
defined selection function and  provides an indication of
cluster richness:  $N_{gals}$. Finally, we checked with the SDSS cluster  
sample, which provides a more robust determination of the cluster richness (i.e. $N_{200}$), 
that $N_{gals}$ in the NSC is a valid measure of the cluster richness, at least for statistical studies like this one. 

\subsection{The $L_{IR}/L_X$  and the dust-to-gas mass ratios}

As shown in Sect.~\ref{ICdust}, we require  a dust-to-gas mass ratio of
$Z_{dust} = 5\times 10^{-4}$ to explain the IR to X-ray luminosity ratio 
if dust in the ICM is the unique source of the IR emission.
This ICM dust abundance is actually to be taken as an upper limit since we obviously know that there is also
a contribution to the IR from the star formation in the member galaxies.
The comparison with the IR-SFR to X-ray luminosity ratios measured for the two clusters
 Coma and MS1054-0321 \citep{Bai07} confirms that the measured SFR may be insufficient to
 explain the high IR to X-ray luminosity ratio derived by our stacking analysis.
 Moreover, our upper limit to the dust in the ICM is of the same order as that measured
 by \citet{Muller08} from  their search for statistical reddening of background galaxies behind a sample of 
 458 RCD clusters with $z < 0.5$. Their extinction value converted into an
upper limit to the dust mass within the central Mpc is $8\times
10^{9}$~M$_\odot$. This value then translates into a dust mass
abundance limit of $Z_{d} \le 2\times 10^{-4}$.
 
Conversely, one can use the existing constraints on the intracluster dust
abundance to derive the $L_{IR}/L_X$ ratio.  By looking at the
reddening of QSOs behind SDSS clusters, \citet{Chelouche07}
measured an average color excess of $<E(g-i)> \simeq 3\times 10^{-3}$~mag.
This detection is valid for the external parts of
the SDSS clusters (i.e. $R > R_{200} \simeq 1$~Mpc). If we extrapolate this
to the central Mpc of a cluster, this 
corresponds to a total dust mass of $M_{dust}=3\times 10^{8}$~M$_\odot$
(see their Eq.~4). Compared with the corresponding gas mass in the same volume,
this leads to $Z_{d} \simeq 10^{-5}$. Note that our stacking integrates the luminosity 
within a disk area of $10'$ radius, which translates into $1$~Mpc radius at $z = 0.1$ and
respectively $4$~Mpc radius at $z = 0.6$. The dust abundance derived from this extinction 
measurement is consistent with the upper limit derived by \citet{Bovy08} from measurement of reddening
of SDSS galaxies behind galaxy clusters and with theoretical predictions given by \citet{Popescu00}. 
From Eq. (\ref{IR2X}), a dust abundance $Z_{d} \simeq 10^{-5}$  translates into $\frac{L_{IR}}{L_X} =
0.4$. This is only about $2 \%$ of the IR luminosity measured in the stacked data, leaving most of it ($\simeq 98\%$) to SFR in the member galaxies.

\subsection{Selection effects}

\begin{figure}[!t]
\resizebox{\hsize}{!}{\includegraphics{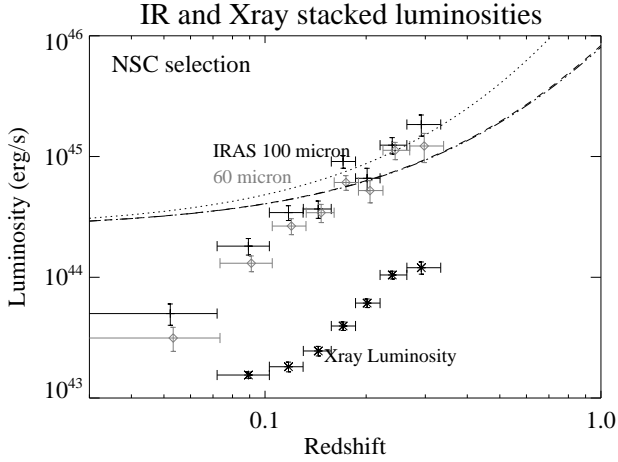}}
\caption{\label{Fig_NSC} Same as Fig.~\ref{Fig_luminozities} but for the NSC catalogue. }
\end{figure}

Obviously the cluster richness for the NED extraction  increases with 
redshift as a result of ill-determined selection effects, and this will affect the stacked luminosities (IR and X-ray). 

\begin{figure}[!t]
\resizebox{\hsize}{!}{\includegraphics{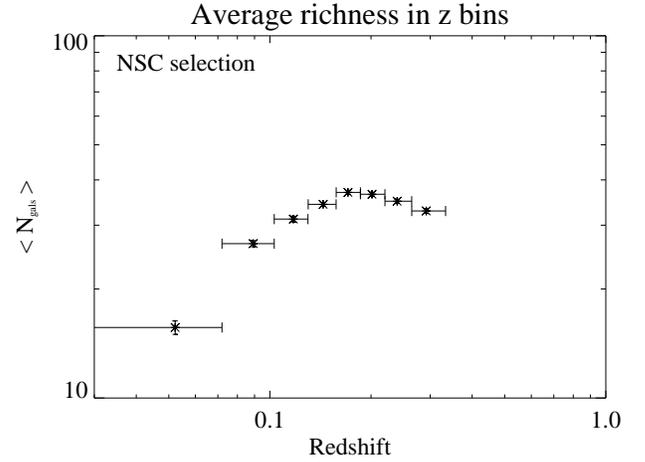}}
\caption{\label{Fig_ngals} Average richness, $<N_{gals}>$, into the redshift bins for the NSC sample. }
\end{figure}

To confirm this, we reproduced the entire stacking analysis for the 
NSC selection. The NSC spans a narrower redshift range than the
NED selection ($z < 0.4$ for the NSC sample, and $z < 1$ for the
NED sample), but it includes an estimation of the cluster richness.  To quantify the selection bias, we   
averaged the cluster richness in the same redshift bins as those fixed for the IR and X-ray luminosities.
The stacked luminosities for the NSC sample are plotted in Fig.~\ref{Fig_NSC}. 
We have retained the same redshift binning as for the NED
analysis, meaning that we now have unequal numbers of clusters in each
bin. We don't show the final two bins which have too few clusters to
provide a statistical significant result (these two bins contain respectively 61 and 14
clusters, whereas we have more than 350 clusters in all other bins).
The average richness in each bin, $< N_{gals}>$, is shown in  Fig. \ref{Fig_ngals}.

The IR luminosities derived from the NSC selection show the same
behaviour as those obtained from the NED selection. The average
richness in the bins indicates that the selection is homogeneous for $z >
0.1$ with a mean NSC richness of $<N_{gals}> \simeq 35$. As expected, there
is a clear bias toward smaller clusters at $z < 0.1$, with a mean
richness of $<N_{gals}> \simeq 16$ for the first bin, corresponding to $z < 0.07$.

\begin{figure}[!t]
\resizebox{\hsize}{!}{\includegraphics{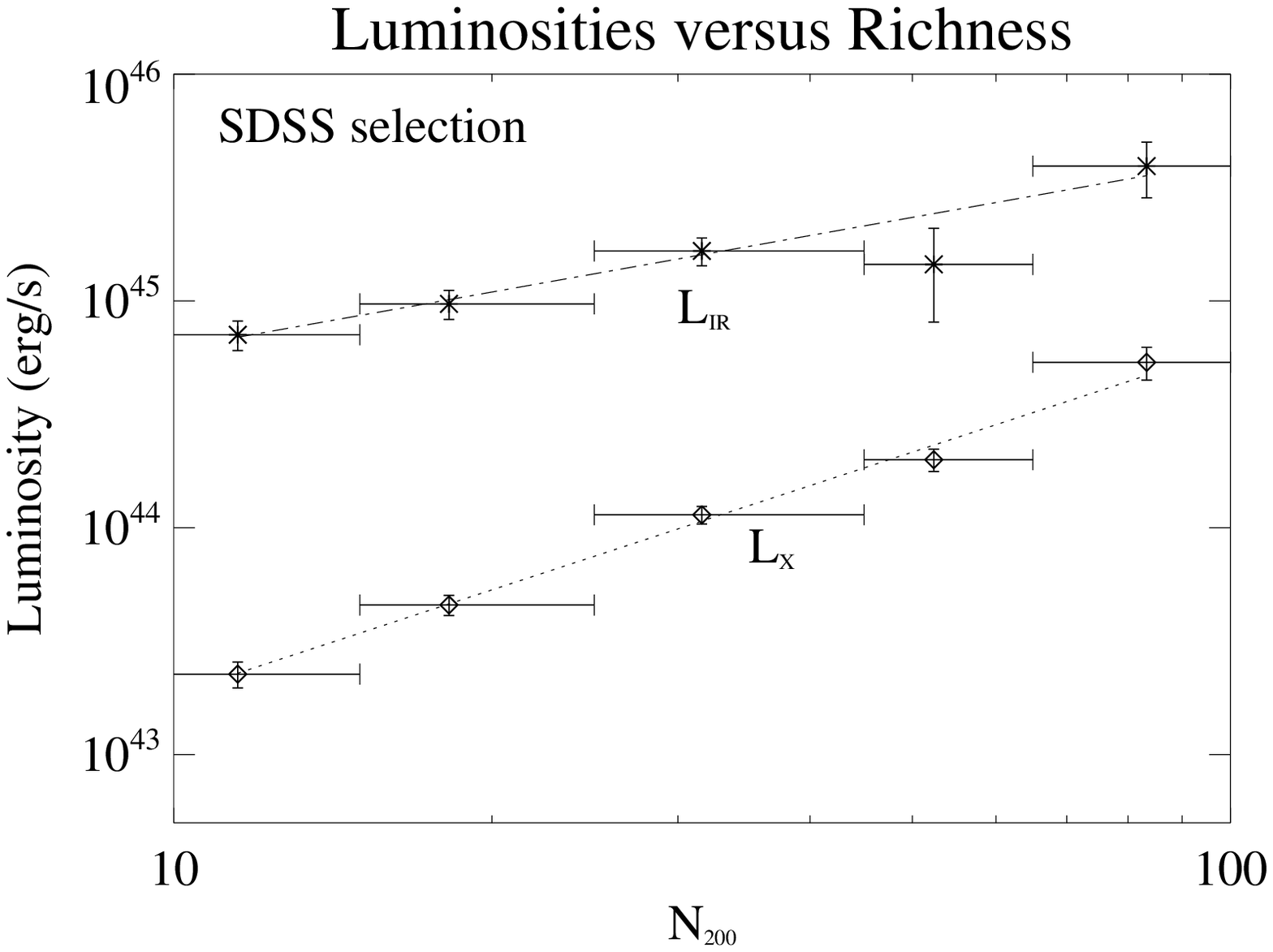}}
\resizebox{\hsize}{!}{\includegraphics{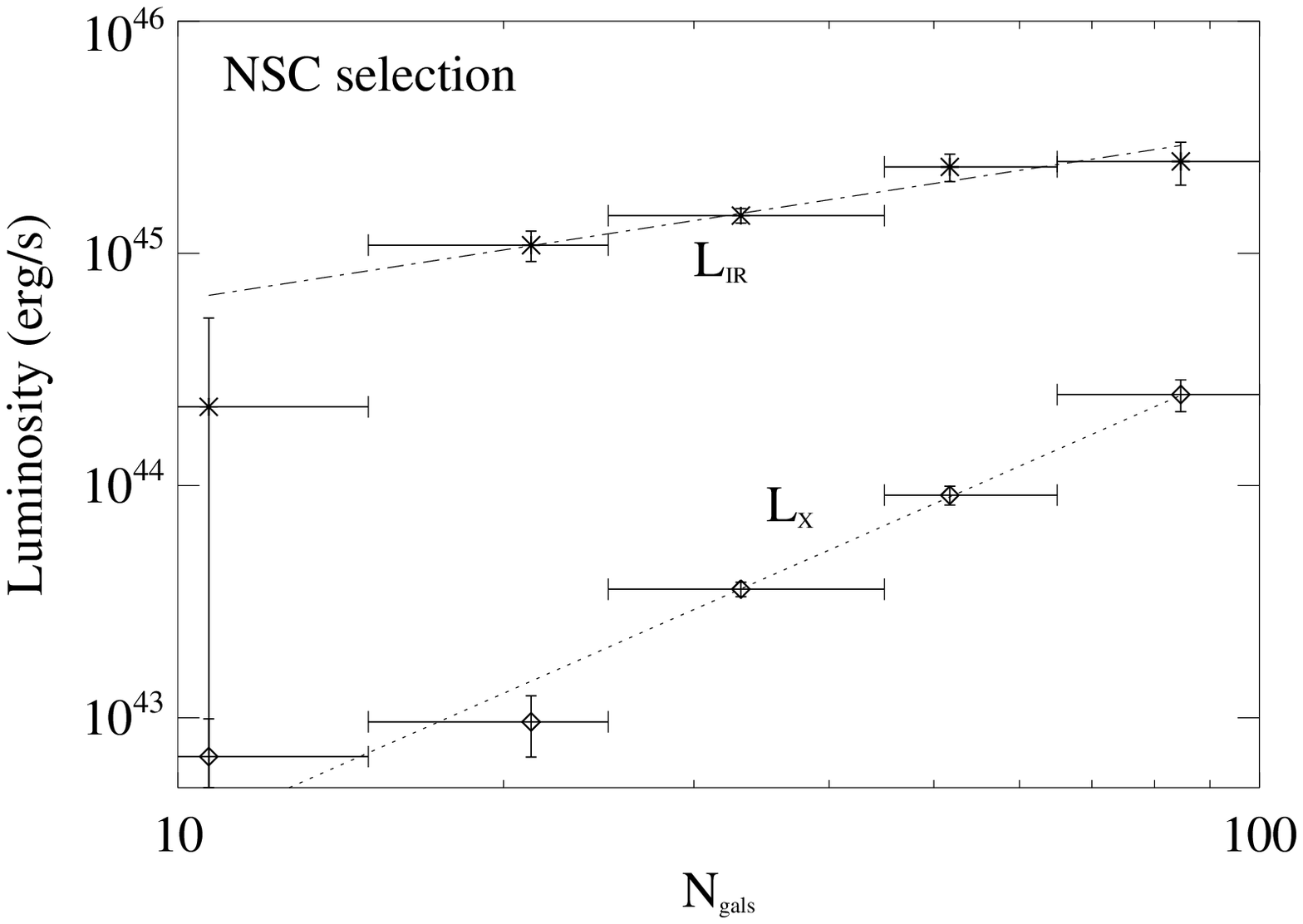}}
\caption{\label{Fig_N200_MaxBCG} IR and X-ray luminosities as a function of the cluster richness for the SDSS sample (upper panel) and the NSC sample (lower panel). The dotted and dashed-dotted lines show the best fit power law functions  describing the data: the exponent is $1.5\pm0.2$ (SDSS) for the X-ray luminosities and $0.83\pm0.2$ for the infrared.
}
\end{figure}

Finally, we checked that the richness parameter  $N_{gals}$ given in the NSC catalogue is a robust probe of the cluster richness. We show in the lower panel of Fig.~\ref{Fig_N200_MaxBCG} that the X-ray luminosity correlates well with $N_{gals}$. To compile this plot, we simply run the stacking again, but for clusters stacked into bins of equal $N_{gals}$ instead of redshift. However, for this analysis, the stacking procedure was modified to compute the background subtraction and luminosity of each cluster field individually, before stacking. The correlation between $N_{gals}$ and the infrared luminosity is good, but not as significant as for the X-ray luminosity, particularly because of the large error bar in the first richness bin. 

From this analysis, we derived the following power laws, correlating the X-ray and infrared luminosities to $N_{gals}$: 

\begin{eqnarray}
L_X  = 10^{42.6\pm0.3} (\frac{N_{gals}}{10})^{2.0\pm0.4} ~ erg/s\nonumber \\
L_{IR} = 10^{44.8\pm0.1} (\frac{N_{gals}}{10})^{0.7\pm0.2} ~ erg/s
\end{eqnarray}

We completed the same analysis for the SDSS selection and the corresponding cluster richness $N_{200}$. The results are plotted in the upper panel of Fig.~\ref{Fig_N200_MaxBCG}. They indicate that both the X-ray and the infrared luminosities correlate well with $N_{200}$. This is not surprising because the richnesses given for the SDSS clusters by \citet{Koester07} were carefully determined in a homogenous way. The correlation analysis provides the following power laws: 

\begin{eqnarray}
L_X  = 10^{43.3\pm0.1} (\frac{N_{200}}{10})^{1.5\pm0.2} ~ erg/s \nonumber \\
L_{IR} = 10^{44.8\pm0.06} (\frac{N_{200}}{10})^{0.8\pm0.2} ~ erg/s
\end{eqnarray}

If we compare our $L_X / N_{200}$ relation with that derived by \citet{Rykoff08} (their Fig.~6), we see that our results  are compatible for $N_{200} \simeq 50$,  given that our luminosities are bolometric and refer to $h = 0.7$ instead of $h = 1$ (i.e. $L_{bol} \simeq 1.35 L_{[0.1-2.4keV]} h^{-2}$). However, their slope is steeper because we did not scale our aperture in terms of either redshift or cluster richness in the same way as these authors ( 750 kpc or $R_{200}$). Therefore we measure slightly higher luminosities for the low richness clusters ($N_{200} < 30$).

The tight correlation between X-ray luminosity and richness is clear both for the NSC and the SDSS clusters and confirms that $N_{gals}$ in the NSC catalogue is a robust measure of the cluster richness, at least for statistical purposes as in Fig. \ref{Fig_ngals} to investigate whether the cluster richness is biased with redshift. 

A secondary result of this analysis is that the exponent of the $L_{IR}$ to $N_{gals}$ correlation is lower than for the X-ray (0.8 instead of 1.5 for the SDSS clusters). This is definitely related to differences in the physical emission processes that occur in the infrared and the X-rays. However, we do not investigate this property in this paper.

\section{Conclusion \label{conclusion}}

We have coadded IRAS and RASS images in the direction of thousands of galaxy clusters selected in the NED database for their low noise properties (smooth background) following the same method as \citet{Montier05}. 
We have compared the IR and X-ray luminosities, and investigated their evolution with redshift. 
We showed that the averaged infrared luminosities are on average 20 times higher than the X-ray luminosities, and that the infrared luminosity increases with redshift approximately as $(1+z)^5$ for redshifts between 0.1 and 1. We applied the same method to clusters selected from the NSC catalogue by \citet{Gal03}. This is  a well selected catalogue of galaxy clusters for which a qualitative control of the selection bias with the redshift was possible. This  showed the same behavior as the previous catalogue, but for a restricted redshift range  $0.1 < z < 0.4$, which confirmed our result for the evolution of the IR luminosity.

If all the infrared luminosity that we measure  is attributed to emission from diffuse intracluster dust, this implies a dust-to-gas mass abundance in the intracluster gas of $Z_d \simeq 5 \times 10^{-4}$. However, given other constraints on the abundance of intracluster dust ($Z_d \simeq 10^{-5}$ from the extinction towards the outer parts of the SDSS clusters measured by \citet{Chelouche07} and \citet{Bovy08}), and the fact that the IR luminosity mimics the evolution of the SFR with redshift, we believe that it is more likely that the main contribution to the cluster IR luminosity is from the ongoing star formation in the member galaxies. From theoretical predictions calibrated with the afore mentioned extinction measurement, we expect that only a minor contribution (a few percent) could originate in intracluster dust. This conclusion is supported independently by the overall shape of the spectral energy distribution of the stacked data that we published in \citet{Montier05}, and which showed a high $12$  to $100~\mu m$ color ratio (of 0.062 for flux density, or 0.52 for luminosity). This high color ratio implies that nanoparticles emitting in the mid IR are abundant relative to big grains within the emitting phase. These tiny particles would not survive in hot intracluster gas, and are instead typical of normal spiral and star-forming galaxies.

\section{Acknowledgments}

We thank the Referee, C. Popescu, for pointing to us an important inconsistency in the former version of the paper and for several suggestions and comments which allowed us to significantly improve the paper.\\
For this work the authors made use of the SIMBAD database, operated at CDS, Strasbourg, France, and of the NASA/IPAC Extragalactic Database (NED) which is operated by the JPL, CALTECH, under contract with NASA. 
EP and LM acknowledge the support of grant ANR-06-JCJC-0141.

\end{document}